\documentclass[11pt]{article}
\usepackage{verbatim,amsmath,amsthm,eufrak}
\usepackage{cite} % prevents showidx from correct work
\usepackage{mathrsfs}
\usepackage{makeidx}
\usepackage{yhmath}
\usepackage{enumitem}
\usepackage{color,xcolor}
\usepackage[hypertexnames=false,hyperfootnotes=false,colorlinks=true,linkcolor=blue,%
citecolor=purple,filecolor=magenta,urlcolor=cyan,unicode,linktocpage=true,pagebackref=false]{hyperref}
\usepackage{scalerel,stackengine}

\def\hhref#1{\href{http://arxiv.org/abs/#1}{arXiv:#1}} % in bibliography
\renewcommand{\arraystretch}{1.2}
\makeatletter
\newdimen\normalarrayskip              % skip between lines
\newdimen\minarrayskip                 % minimal skip between lines
\normalarrayskip\baselineskip
\minarrayskip\jot
\newif\ifold             \oldtrue            \def\new{\oldfalse}
\def\arraymode{\ifold\relax\else\displaystyle\fi} % mode of array entries
\def\eqnumphantom{\phantom{(\theequation)}}     % right phantom in eqnarray
\def\@arrayskip{\ifold\baselineskip\z@\lineskip\z@
     \else
     \baselineskip\minarrayskip\lineskip2\minarrayskip\fi}
\def\@arrayclassz{\ifcase \@lastchclass \@acolampacol \or
\@ampacol \or \or \or \@addamp \or
   \@acolampacol \or \@firstampfalse \@acol \fi
\edef\@preamble{\@preamble
  \ifcase \@chnum
     \hfil$\relax\arraymode\@sharp$\hfil
     \or $\relax\arraymode\@sharp$\hfil
     \or \hfil$\relax\arraymode\@sharp$\fi}}
\def\@array[#1]#2{\setbox\@arstrutbox=\hbox{\vrule
     height\arraystretch \ht\strutbox
     depth\arraystretch \dp\strutbox
     width\z@}\@mkpream{#2}\edef\@preamble{\halign
\noexpand\@halignto
\bgroup \tabskip\z@ \@arstrut \@preamble \tabskip\z@ \cr}%
\let\@startpbox\@@startpbox \let\@endpbox\@@endpbox
  \if #1t\vtop \else \if#1b\vbox \else \vcenter \fi\fi
  \bgroup \let\par\relax
  \let\@sharp##\let\protect\relax
  \@arrayskip\@preamble}
\def\eqnarray{\stepcounter{equation}%
              \let\@currentlabel=\theequation
              \global\@eqnswtrue
              \global\@eqcnt\z@
              \tabskip\@centering
              \let\\=\@eqncr
 \halign to \displaywidth\bgroup
    \eqnumphantom\@eqnsel\hskip\@centering
    $\displaystyle \tabskip\z@ {##}$%
    \global\@eqcnt\@ne \hskip 2\arraycolsep
         $\displaystyle\arraymode{##}$\hfil
    \global\@eqcnt\tw@ \hskip 2\arraycolsep
         $\displaystyle\tabskip\z@{##}$\hfil
         \tabskip\@centering
    &{##}\tabskip\z@\cr}
\begingroup\ifx\undefined\newsymbol \else\def\input#1 {\endgroup}\fi

\newcounter{app}

\def\app{\setcounter{equation}{0}
\def\theequation{A\Roman{app}.\arabic{equation}}\par
   \addvspace{4ex}
   \@afterindentfalse
  \secdef\@app\@dapp}
\newcommand\@app{\@startsection {app}{1}{0ex}%
                                   {-3.5ex \@plus -1ex \@minus -.2ex}%
                                   {2.3ex \@plus.2ex}%
                                   {\normalfont\Large\bf}}

\def\@dapp#1{%
{\parindent \z@ \raggedright  \bf #1}\par\nobreak}
\def\l@app#1#2{\ifnum \c@tocdepth >\z@
    \addpenalty\@secpenalty
    \addvspace{1.0em \@plus\p@}%
    \setlength\@tempdima{8.5em}%
    \begingroup
      \parindent \z@ \rightskip \@pnumwidth
      \parfillskip -\@pnumwidth
      \leavevmode \bfseries
      \advance\leftskip\@tempdima
      \hskip -\leftskip
      #1\nobreak\hfil \nobreak\hb@xt@\@pnumwidth{\hss #2}\par
    \endgroup\fi}
\newcounter{sapp}[app]

\def\sapp{\def\theequation{A\arabic{app}.\arabic{equation}}\par
   \@afterindentfalse
  \secdef\@sapp\@dsapp}
\newcommand\@sapp{\@startsection{sapp}{2}{\z@}%
                                     {-3.25ex\@plus -1ex \@minus -.2ex}%
                                     {1.5ex \@plus .2ex}%
                                     {\normalfont\large\bfseries}}

\def\@dsapp#1{%
{\parindent \z@ \raggedright  \bf #1}\par\nobreak}
\newcommand{\l@sapp}{\@dottedtocline{2}{1.5em}{3em}}
\def\draft{\oddsidemargin -.5truein
        \def\@oddfoot{\sl preliminary draft \hfil
        \rm\thepage\hfil\sl\today\quad\militarytime}
        \let\@evenfoot\@oddfoot \overfullrule 3pt
        \let\label=\draftlabel
        \let\marginnote=\draftmarginnote
   \def\@eqnnum{(\theequation)\rlap{\kern\marginparsep\tt\@eqnlabel}%
\global\let\@eqnlabel\@vacuum}  }

\def\be{\begin{eqnarray}}
\def\ee{\end{eqnarray}}

\def\p{\partial}

\def\beq{\begin{equation}}
\def\eeq{\end{equation}}
\def\ba{\beq\new\begin{array}{c}}
\def\ea{\end{array}\eeq}
\def\be{\ba}
\def\ee{\ea}

\def\Tr{{\rm Tr}\,}

\def\diag{{\rm diag}\,}

\newfont{\Bbbb}{msbm7 scaled 1\@ptsize00}

\newcommand{\z}{\raise-1pt\hbox{$\mbox{\Bbbb Z}$}}
\newcommand{\n}{\raise-1pt\hbox{$\mbox{\Bbbb N}$}}
\newcommand{\rr}{\raise-1pt\hbox{$\mbox{\Bbbb R}$}}
\newcommand{\cc}{\raise-1pt\hbox{$\mbox{\Bbbb C}$}}

\def\gl{{\rm gl}}

\newcommand{\<}{\left <}
\renewcommand{\>}{\right >}

\newfont{\alef}{msbm10 at 11pt}
\newfont {\goth}{eufm10 at 11pt}
\def\mathbb#1{\hbox{{\alef #1}}}

\DeclareMathOperator{\DP}{DP}
\DeclareMathOperator{\GL}{GL}

\DeclareMathOperator{\odd}{odd}

\let\@@savethanks\thanks
\def\thanks#1{\gdef\thefootnote{\alph{footnote}}\@@savethanks{#1}}

\newtheorem{theorem}{Theorem}%[section]
\newtheorem{lemma}{Lemma}[section]

\newtheorem{corollary}[lemma]{Corollary}
\newtheorem{remark}{Remark}[section]

\newtheorem{conjecture}{Conjecture}%[section]
\newtheorem*{theorem*}{Theorem}

\numberwithin{equation}{section}

\makeatletter
\g@addto@macro \normalsize {%
 \setlength\abovedisplayskip{14pt plus 3pt minus 3pt}%
 \setlength\belowdisplayskip{14pt plus 3pt minus 3pt}%
  \setlength\abovedisplayshortskip{11pt plus 3pt minus 3pt}%
 \setlength\belowdisplayshortskip{11pt plus 3pt minus 3pt}%
}
\makeatother

\bigskip
\bigskip

\title{
\bigskip
{\bf Intersection numbers on  $\overline {\mathcal M}_{g,n}$ and BKP hierarchy} \vspace{.5cm}}
\author{{\bf Alexander Alexandrov}\thanks{E-mail:  {\tt alexandrovsash at gmail.com}}
\date{ } \\
{\small {\it Center for Geometry and Physics, Institute for Basic Science (IBS), Pohang 37673, Korea  }}
}

\begin{document}

\setcounter{footnote}{0}

\setcounter{tocdepth}{3}

\maketitle

\vspace{-8.0cm}

\begin{center}
%\hfill ITEP/TH-18/16
\end{center}

\vspace{6.5cm}

\begin{center}
{\em Dedicated to the memory of Sergey Mironovich Natanzon}
\end{center}

\vspace{0.5cm}

\begin{small}\begin{center}
\today
\end{center}\end{small}
%\bigskip
\begin{abstract} 
In their recent inspiring paper Mironov and Morozov claim a surprisingly simple expansion formula for the Kontsevich--Witten tau-function
in terms of the Schur Q-functions. Here we provide a similar conjecture for the Br\'ezin--Gross--Witten tau-function. Moreover, we identify both tau-functions of the KdV hierarchy, which describe intersection numbers on the moduli spaces of punctured Riemann surfaces, with the hypergeometric solutions of the BKP hierarchy.
\end{abstract}
\bigskip

%{Keywords: enumerative geometry, matrix models, tau-functions, KP hierarchy, Virasoro constraints, cut-and-join operator}\\

\bigskip

{\small \bf MSC 2020 Primary: 37K10, 14N35, 81R10, 14N10, 05A15; Secondary:  81T32.}

\begin{comment}
(	
37K10 Completely integrable infinite-dimensional Hamiltonian and Lagrangian systems, integration methods, integrability tests, integrable hierarchies (KdV, KP, Toda, etc.)
14N35 Gromov-Witten invariants, quantum cohomology, Gopakumar-Vafa invariants, Donaldson-Thomas invariants (algebro-geometric aspects) 
81R10 Infinite-dimensional groups and algebras motivated by physics, including Virasoro, Kac-Moody, W-algebras
and other current algebras and their representations 
81R12 Groups and algebras in quantum theory and relations with integrable systems
14H70 Relationships between algebraic curves and integrable systems
81T32 Matrix models and tensor models for quantum field theory
05A15 Exact enumeration problems, generating functions
14N10 Enumerative problems (combinatorial problems) in algebraic geometry
15B52 Random matrices (algebraic aspects) 
81T40 Two-dimensional field theories, conformal field theories, etc. in quantum mechanics
\end{comment}

%\bigskip

%\bigskip

\newpage

%\tableofcontents

\def\thefootnote{\arabic{footnote}}
\section{Introduction}
%\addcontentsline{toc}{section}{Introduction}
%\def\theequation{\arabic{equation}}
\setcounter{equation}{0}

Kontsevich--Witten (KW) tau-function \cite{Kon92,Wit91} of the KdV hierarchy plays a special role in modern mathematical physics. It is the basic building block for several universal constructions, including Chekhov--Eynard--Orantin topological recursion and Givental decomposition. This makes the Kontsevich--Witten tau-function one of the most well-studied tau-functions of the integrable solitonic hierarchies. However, unexpected new properties of this tau-function continue to surprise.

In their new paper \cite{MMQ} Mironov and Morozov came with a new idea. According to their conjecture, expansion of the Kontsevich--Witten tau-function in the basis of the Schur Q-functions $Q_\lambda$ is unexpectedly simple: 
\be
\tau_{KW}= \sum_{\lambda \in \DP} \left(\frac{\hbar}{16}\right)^{|\lambda|/3} \frac{Q_\lambda({\bf t}) Q_\lambda(\delta_{k,1})Q_{2\lambda}(\delta_{k,3}/3)}{Q_{2\lambda}(\delta_{k,1})}.
\ee
Here the summation runs over strict partitions.

The Br\'ezin--Gross--Witten (BGW) model was introduced in the lattice gauge theory 40 years ago \cite{Brezin, GW}. 
It has a natural enumerative geometry interpretation given by the intersection theory of Norbury's $\Theta$-classes, also related to super Riemann surfaces \cite{Norb,N2}. Moreover, the BGW tau-function is another fundamental element of  topological recursion/Givental decomposition, which corresponds to the hard edge case \cite{AMMBGW,CN}. Similarly to the KW case, it is a tau-function of the KdV integrable hierarchy and can be described by the generalized Kontsevich model \cite{MMS}. All this makes the BGW model interesting and, in many respects, similar to the KW tau-function. 

In this paper we conjecture a Schur Q-function expansion of the BGW tau-function. Similarly to the Mironov--Morozov formula for the KW tau-function, this expansion is described by simple coefficients made of specifications of the Schur Q-functions
\be
\tau_{BGW}= \sum_{\lambda \in \DP} \left(\frac{\hbar}{16}\right)^{|\lambda|} \frac{Q_\lambda({\bf t}) Q_\lambda(\delta_{k,1})^3}{Q_{2\lambda}(\delta_{k,1})^2}.
\ee

These expansions leads to the question about the relation between KW and BGW tau-functions on one side, and BKP hierarchy on another. Indeed, Schur Q-functions are known to provide a natural basis for the expansion of the BKP tau-functions.
Surprisingly enough,  we found that KW and generalized BGW tau-functions after a simple rescaling of the times satisfy the BKP integrable hierarchy. Moreover, they belong to the family of the {\em hypergeometric} BKP tau-functions, related to {\em spin Hurwitz numbers}.

While a non-linear relation between tau-functions of the KP and BKP hierarchies is well-known \cite{Date}, the identification of KdV and BKP tau-functions is new and unexpected (see, however, \cite{ASY}). In this paper we investigate only a few examples of such a relation.
It would be interesting to find a general intersection of the solutions of KdV and BKP hierarchies
\be
{\rm KdV}\cap {\rm BKP}=?,
\ee
or, equivalently, all functions $\tau$ such that $\tau({\bf t})$ is a tau-function of the KdV hierarchy and $\tau({\bf t}/2)$ is a tau-function of the BKP hierarchy. Interpretation of this class of tau-functions in terms of usual and orthogonal Sato Grassmannians is rather intriguing.

The present paper is organized as follows. In Section \ref{S1} we remind the reader some basic elements of the intersection theory on the moduli spaces and its relation to the Kontsevich--Witten and Br\'ezin--Gross--Witten tau-functions. Section \ref{S2} is devoted to Schur Q-function expansion of these tau-functions. In Section \ref{S3} we describe interpretation of these tau-functions as the hypergeometric solutions of the BKP hierarchy.

%%%%%%%%%%%%%%%%%%%%%%%%%%%%%%%%%%%%%%%%%%%

\section{KW and BGW tau-functions in intersection theory}\label{S1}

%%%%%%%%%%%%%%%%%%%%%%%%%%%%%%%%%%%%%%%%%%%

\subsection{Intersection numbers and their generating functions}

Denote by $\overline {\mathcal M}_{g,n}$ the Deligne--Mumford compactification of the moduli space of all compact Riemann surfaces of genus~$g$ with~$n$ distinct marked points. It is a non-singular complex orbifold of dimension~$3g-3+n$. It is empty unless the stability condition
\begin{gather}\label{stability}
2g-2+n>0
\end{gather}
is satisfied. 

New directions in the study of $\overline{\mathcal{M}}_{g,n}$ were initiated by Witten in his seminal paper~\cite{Wit91}. For each marking index~$i$ consider the cotangent line bundle ${\mathbb{L}}_i \rightarrow \overline{\mathcal{M}}_{g,n}$, whose fiber over a point $[\Sigma,z_1,\ldots,z_n]\in \overline{\mathcal{M}}_{g,n}$ is the complex cotangent space $T_{z_i}^*\Sigma$ of $\Sigma$ at $z_i$. Let $\psi_i\in H^2(\overline{\mathcal{M}}_{g,n},\mathbb{Q})$ denote the first Chern class of ${\mathbb{L}}_i$. We consider the intersection numbers
\begin{gather}\label{eq:products}
\<\tau_{a_1} \tau_{a_2} \cdots \tau_{a_n}\>_g:=\int_{\overline{\mathcal{M}}_{g,n}} \psi_1^{a_1} \psi_2^{a_2} \cdots \psi_n^{a_n}.
\end{gather}
The integral on the right-hand side of~\eqref{eq:products} vanishes unless the stability condition~\eqref{stability} is satisfied, all  $a_i$ are non-negative integers, and the dimension constraint 
\be\label{d1}
3g-3+n=\sum_{i=1}^n a_i
\ee 
holds true. Let $T_i$, $i\geq 0$, be formal variables and let
\be
\tau_{KW}:=\exp\left(\sum_{g=0}^\infty \sum_{n=0}^\infty \hbar^{2g-2+n}F_{g,n}\right),
\ee
where
\be
F_{g,n}:=\sum_{a_1,\ldots,a_n\ge 0}\<\tau_{a_1}\tau_{a_2}\cdots\tau_{a_n}\>_g\frac{\prod T_{a_i}}{n!}.
\ee
Witten's conjecture \cite{Wit91}, proved by Kontsevich \cite{Kon92}, states that the partition function~$\tau_{KW}$ becomes a tau-function of the KdV hierarchy after the change of variables~$T_n=(2n+1)!!t_{2n+1}$.
% Integrability immediately follows from Kontsevich's matrix integral representation, for more details see \cite{H3_2} and references therein.

On the same moduli space one can consider other types of intersection numbers. An interesting family of such intersection numbers was recently considered by  Norbury \cite{Norb}.
Namely, he introduced $\Theta$-classes, $\Theta_{g,n}\in H^{4g-4+2n}(\overline{\mathcal{M}}_{g,n})$, and described their intersections with the $\psi$-classes
\be
\<\tau_{a_1} \tau_{a_2} \cdots \tau_{a_n}\>_g^\Theta= \int_{\overline{\mathcal{M}}_{g,n}}\Theta_{g,n} \psi_1^{a_1} \psi_2^{a_2} \cdots \psi_n^{a_n}. 
\ee
Again, the integral on the right-hand side vanishes unless the stability condition~\eqref{stability} is satisfied, all $a_i$ are non-negative integers, and the dimension constraint 
\be\label{d2}
g-1=\sum_{i=1}^n a_i
\ee 
holds true.
Consider the generating function of the 
intersection numbers of $\Theta$-classes and $\psi$-classes
\be
F^\Theta_{g,n}= \sum_{a_1,\ldots,a_n\ge 0}  \frac{\prod T_{a_i}}{n!} \int_{\overline{\mathcal{M}}_{g,n}}\Theta_{g,n} \psi_1^{a_1} \psi_2^{a_2} \cdots \psi_n^{a_n} 
\ee
then, we have a direct analog of the Kontsevich--Witten tau-function \cite{Norb}:
\begin{theorem*}[Norbury]
Generating function
\be
\tau_\Theta = \exp\left(\sum_{g=0}^\infty \sum_{n=0}^\infty \hbar^{2g-2+n}F_{g,n}^\Theta \right)
\ee
becomes a tau-function of the KdV hierarchy after the change of variables~$T_n=(2n+1)!!t_{2n+1}$.
\end{theorem*}
Norbury also proved, that $\tau_\Theta$ is nothing but a tau-function of the Br\'ezin--Gross--Witten model \cite{GW,Brezin}
\be
\tau_\Theta=\tau_{BGW}.
\ee
We refer the reader to \cite{Norb,N2} for a detailed description. 

Both KW and BGW tau-functions can be described by matrix models. Consider a diagonal matrix $\Lambda=\diag(\lambda_1,\lambda_2,\dots,\lambda_M)$.
For any function $f$, dependent on the infinite set of variables ${\bf t}=(t_1,t_2,t_3,\dots)$, let
\be\label{Miwa}
f\left(\left[\Lambda^{-1}\right]\right):=f({\bf t})\Big|_{t_k=\frac{1}{k}\Tr \Lambda^{-k}}
\ee
be the {\em Miwa parametrization}. The KW tau-function can be described by the Kontsevich matrix integral \cite{Kon92} 
\be\label{mati}
\tau_{KW}([\Lambda^{-1}]):={\mathcal C}^{-1} \int[ d\Phi] \exp\left(-\frac{1}{\hbar}\Tr\left(\frac{\Phi^3}{3!}+\frac{\Lambda\Phi^2}{2} \right)\right).
\ee
Here one takes the asymptotic expansion of the  integral over Hermitian $M\times M$ matrices. 
The BGW tau-function in the Miwa parametrization is given by the unitary matrix integral
\be
Z_{BGW}([(A^\dagger A)^{-\frac{1}{2}}])=\int \left[d U\right] e^{\frac{1}{2\hbar}\Tr ( A^\dagger U+A U^\dagger)}.
\ee
It also has another integral description \cite{MMS}, similar to the Kontsevich integral (\ref{mati}), see Section \ref{GBGW} below. KdV integrability of the BGW model easily follows from this description.
%%%%%%%%%%%%%%%%%%%%%%%%%%%%%%%%%%%%%%%%%%%

\subsection{Heisenberg--Virasoro constraints and cut-and-join description}

The KW and BGW tau-functions are solutions of the KdV hierarchy, which is a reduction of the KP hierarchy. In terms of tau-function $\tau({\bf t})$ it is described by the Hirota bilinear identity
\be\label{HBEKP}
\oint_{\infty} e^{\xi({\bf t-t'},z)}
\tau ({\bf t}-[z^{-1}])\tau ({\bf t'}+[z^{-1}])dz =0.
\ee
This bilinear identity encodes all nonlinear equations of the KP hierarchy. 
Here we use the standard short-hand notations
\be
{\bf t}\pm [z^{-1}]:= \bigl \{ t_1\pm   
z^{-1}, t_2\pm \frac{1}{2}z^{-2}, 
t_3 \pm \frac{1}{3}z^{-3}, \ldots \bigr \}
\ee
and
\be
\xi({\bf t},z)=\sum_{k>0} t_k z^k.
\ee
If a tau-function $\tau({\bf t})$ of the KP hierarchy does not depend on even time variables,
\be\label{Red}
\frac{\p}{\p t_{2k}}\tau({\bf t})=0 \,\,\,\,\,\,\,\,\,\,\,\, \forall k>0,
\ee 
than it is a tau-function of the KdV hierarchy.

Symmetries of the KP hierarchy can be described in terms of a central extension of the $\GL(\infty)$ group. Let us consider
the {\em Heisenberg--Virasoro subalgebra} of the corresponding $\gl(\infty)$ algebra. It is generated by the operators
\be
\widehat{J}_k =
\begin{cases}
\displaystyle{\frac{\p}{\p t_k} \,\,\,\,\,\,\,\,\,\,\,\, \mathrm{for} \quad k>0},\\[2pt]
\displaystyle{0}\,\,\,\,\,\,\,\,\,\,\,\,\,\,\,\,\,\,\, \mathrm{for} \quad k=0,\\[2pt]
\displaystyle{-kt_{-k} \,\,\,\,\,\mathrm{for} \quad k<0,}
\end{cases}
\ee
unit, and
\be
\label{virfull}
\widehat{L}_m=\frac{1}{2} \sum_{a+b=-m}a b t_a t_b+ \sum_{k=1}^\infty k t_k \frac{\p}{\p t_{k+m}}+\frac{1}{2} \sum_{a+b=m} \frac{\p^2}{\p t_a \p t_b}.
\ee

The KW and BGW tau-functions can be naturally described in terms of this Heisenberg--Virasoro algebra.
Let us introduce the notation 
\be
\tau_1={\tau}_{KW}, \,\,\,\,\,\, \tau_0={\tau}_{BGW}.
\ee
Below we assume that $\alpha\in \{0,1\}$. Both KW and BGW tau-functions are solutions of the KdV hierarchy, hence they satisfy the Heisenberg constraints
\be\label{Heis}
\frac{\p}{\p t_{2k}} {\tau}_\alpha=0, \,\,\,\,\,\,\,\,\,\,\,\, k>0.
\ee
The dimensional constrains (\ref{d1}) and (\ref{d2}) can be represented as
\begin{equation}
\begin{split}\label{dimV}
\widehat{L}_0 \cdot {\tau}_\alpha = (1+2\alpha) \hbar \frac{\p}{\p \hbar}  {\tau}_\alpha.
%\widehat{L}_0 \cdot {\tau}_{BGW} = \hbar \frac{\p}{\p \hbar}  {\tau}_{BGW}. 
\end{split}
\end{equation}
Moreover, these tau-functions satisfy the Virasoro constraints
\begin{equation}
\begin{split}\label{VirC}
\widehat{L}_k^{\alpha} \cdot {\tau}_{\alpha} &=0, \,\,\,\,\,\,\,\,\, k \geq -\alpha, \\
%\widehat{L}_k^{BGW} \cdot {\tau}_{BGW} &=0, \,\,\,\, k \geq -0,
\end{split}
\end{equation}
where the Virasoro operators are given by
\begin{equation}
\begin{split}
\widehat{L}_k^{\alpha}&=\frac{1}{2}\widehat{L}_{2k}-\frac{1}{2\hbar}\frac{\p}{\p t_{2k+1+2\alpha}} +\frac{\delta_{k,0}}{16}.
%\widehat{L}_k^{BGW} &=\frac{1}{2}\widehat{L}_{2k}-\frac{1}{2\hbar}\frac{\p}{\p t_{2k+1}} +\frac{\delta_{k,0}}{16}.
\end{split}
\end{equation}
%These constraints follow from the theorems of Kontsevich and Norbury and earlier results of \cite{GN,Fuku,DVV}.

Following the ideas of \cite{MSh} and combining (\ref{dimV}) with (\ref{Heis}) and (\ref{VirC}) one gets  \cite{ABGW,AKW}
\be\label{caje}
\frac{\p}{\p \hbar} \tau_\alpha=\widehat{W}_\alpha \cdot \tau_\alpha, 
\ee
where
\be
\widehat{W}_\alpha=\frac{1}{2\alpha+1} \sum_{k=0}^\infty (2k+1) t_{2k+1}\left(\widehat{L}_{2k-2\alpha}+\frac{\delta_{k,\alpha}}{8}\right).
\ee
Operators $\widehat{W}_\alpha$ are called {\em cut-and-join} operators because of their similarity to the cut-and-join operator description of simple Hurwitz numbers \cite{caj,caj1}. Below we will work with the space of odd times only, therefore we can represent these operators as follows
\begin{equation}\label{CAJ}
\begin{split}
\widehat{W}_0&=\sum_{k,m\in \z_{\odd}^+}^\infty \left(kmt_{k}t_{m}\frac{\p}{\p t_{k+m-1}}+\frac{1}{2}(k+m+1)t_{k+m+1}\frac{\p^2}{\p t_k \p t_m}\right)+\frac{t_1}{8},\\
\widehat{W}_1&=\frac{1}{3}\sum_{k,m\in \z_{\odd}^+}^\infty \left(kmt_{k}t_{m}\frac{\p}{\p t_{k+m-3}}+\frac{1}{2}(k+m+3)t_{k+m+3}\frac{\p^2}{\p t_k \p t_m}\right)+\frac{t_1^3}{3!}+\frac{t_3}{8},
\end{split}
\end{equation}
where we denote by ${\mathbb Z}^+_{\odd}$ the set of all positive odd integers.
Equation (\ref{caje}) leads to the cut-and-join description for the KW and BGW tau-functions \cite{AKW,ABGW}
\be\label{cajkdv}
\tau_{\alpha}=e^{\hbar \widehat{W}_{\alpha}}\cdot 1.
\ee
This description is convenient for the perturbative computations, for example
%Please note that in \cite{AKW,ABGW} we have used another normalization. 
%where
%\begin{multline}
%\widehat{W}_{BGW}=\frac{1}{2}\sum_{k,m=0}^\infty (2k+1)(2m+1)t_{2k+1}t_{2m+1}\frac{\p}{\p t_{2k+2m+1}}\\
%+\frac{1}{4}\sum_{k,m=0}^\infty(2k+2m+3)t_{2k+2m+3}\frac{\p^2}{\p t_{2k+1}\p t_{2m+1}}+\frac{t_1}{16}
%\end{multline}
%and
%\begin{multline}
%\widehat{W}_{KW}=\frac{1}{3}\sum_{\substack{k,m\geq 0}}\left(2k+1\right)\left(2m+1\right)t_{2k+1}t_{2m+1}\frac{\p}{\p t_{2k+2m-1}}\\
%+\frac{1}{3!}\sum_{k,m\geq 0}\left(2k+2m+5\right)
%t_{2k+2m+5}\frac{\p^2}{\p t_{2k+1}\p t_{2m+1}}+\frac{t_1^3}{3!}+\frac{t_3}{8}.
%\end{multline}
\begin{multline}
\log \tau_0={\frac {t_{{1}}}{8}}\hbar +{\frac {{t_{{1}}}^{2}}{16}}{\hbar }^{2}+
 \left( {\frac {{t_{{1}}}^{3}}{24}}+{\frac {9\,t_{{3}}}{128}} \right) 
{\hbar }^{3}+ \left( {\frac {{t_{{1}}}^{4}}{32}}+{\frac {27\,t_{{3}}t_
{{1}}}{128}} \right) {\hbar }^{4}+ \left( {\frac {{t_{{1}}}^{5}}{40}}+
{\frac {27\,t_{{3}}{t_{{1}}}^{2}}{64}}+{\frac {225\,t_{{5}}}{1024}}
 \right) {\hbar }^{5}\\
 + \left( {\frac {{t_{{1}}}^{6}}{48}}+{\frac {45\,
t_{{3}}{t_{{1}}}^{3}}{64}}+{\frac {1125\,t_{{5}}t_{{1}}}{1024}}+{
\frac {567\,{t_{{3}}}^{2}}{1024}} \right) {\hbar }^{6}+O(\hbar^7).
\end{multline}
\begin{multline}
\log \tau_1=\left( {\frac {{t_{{1}}}^{3}}{6}}+{\frac {t_{{3}}}{8}} \right) 
\hbar + \left( {\frac {t_{{3}}{t_{{1}}}^{3}}{2}}+{\frac {5\,t_{{5}}t_{
{1}}}{8}}+{\frac {3\,{t_{{3}}}^{2}}{16}} \right) {\hbar }^{2}\\
+ \left( 
{\frac {15\,t_{{1}}t_{{3}}t_{{5}}}{4}}+{\frac {3\,{t_{{3}}}^{2}{t_{{1}
}}^{3}}{2}}+{\frac {5\,t_{{5}}{t_{{1}}}^{4}}{8}}+{\frac {35\,t_{{7}}{t
_{{1}}}^{2}}{16}}+{\frac {3\,{t_{{3}}}^{3}}{8}}+{\frac {105\,t_{{9}}}{
128}} \right) {\hbar }^{3}+O(\hbar^4),
\end{multline}

Relation of these cut-and-join operators to the KdV integrability is not known yet. However, operator $\widehat{W}_{0}$ has a natural interpretation in terms of another integrable system, namely the BKP hierarchy, see Section \ref{BKPS} below. 

%%%%%%%%%%%%%%%%%%%%%%%%%%%%%%%%%%%%%%%%%%%

\subsection{Generalized Br\'ezin--Gross--Witten model}\label{GBGW}

{\em Generalized Br\'ezin--Gross--Witten model} was introduced in \cite{MMS} and further investigated in \cite{ABGW}. In the Miwa parametrization it is given by the asymptotic expansion of the matrix integral
\be
\tau_{BGW}([\Lambda^{-1}],N):=\tilde{\mathcal C}^{-1} \int[ d\Phi] \exp\left(-\frac{1}{2\hbar}\Tr\left(\Lambda^2 \Phi+\Phi^{-1}+2\hbar(N-M)\log \Phi\right)\right).
\ee
For $N=0$ it reduces to the original BGW model, $\tau_{BGW}({\bf t},0)=\tau_{BGW}({\bf t})$.
For arbitrary $N$, this is a tau-function of the KdV hierarchy. Moreover, the tau-functions for different values of the parameter $N$ (do not confuse with $M$, the size of the matrices) are related to each other by the MKP hierarchy.
The Virasoro constrains for the generalized BGW model can be derived with the help of the Kac--Schwarz approach \cite{ABGW}. It leads to the cut-and-join description
\be\label{cajg}
\tau_{BGW}({\bf t},N)=e^{\hbar \widehat{W}_0(N)}\cdot 1,
\ee
with the cut-and-join operator given by a deformation of $\widehat{W}_0$ in (\ref{CAJ}),
\be\label{CAJN}
\widehat{W}_0(N)=\widehat{W}_0-\frac{N^2}{2}t_1.
\ee
We expect, that for arbitrary $N$ this generalized model has interesting enumerative geometry interpretation, related to Norbury's $\Theta$-classes.

%%%%%%%%%%%%%%%%%%%%%%%%%%%%%%%%%%%%%%%%%%%

\section{Q-Schur expansion of the KW and BGW tau-functions}\label{S2}

Schur Q-functions were introduced by Schur \cite{SchQ} for the description of the projective representations of the symmetric groups. 
These functions are labeled by {\em strict partitions}. A partition $\lambda$ is strict, if $\lambda_1>\lambda_2>\lambda_3>\dots>\lambda_{\ell(\lambda)}>\lambda_{\ell(\lambda)+1}=0$. We denote the set of strict partitions, including the empty one, by $\DP$.

For the Schur Q-functions we use the same normalization as in \cite{MMQ,MMNQ}. It is related to one, considered by Macdonald in Section 3.8 of his book \cite{Mac}, by 
\be
Q_\lambda= 2^{-\ell(\lambda)/2} \, Q_\lambda^{{\rm{Mac}}}.
\ee
%For the strict partitions they are related to the Schur functions by
%\be
%Q_\lambda({\bf t}) =\left.\sqrt{s_{2\lambda}(\bf t)}\right|_{t_{2k}=0}.
%\ee
In many aspects the Schur Q-functions are similar to the usual Schur functions \cite{Mac}.
For example, let us mention the {\em Cauchy formula}
\be
 \sum_{\lambda \in \DP}  Q_\lambda({\bf t}) Q_\lambda({\bf t'})=\exp \left(2 \sum_{k\in {\z}_{\odd}^+} k t_{k} {t'}_{k}\right),
\ee
and an analog of the standard hook formula \cite{SchQ}
\be\label{hook}
Q_\lambda(\delta_{k,1})=2^{|\lambda|-\ell(\lambda)/2}\frac{1}{\prod_{j=1}^{\ell(\lambda)} \lambda_j!} \prod_{k<m}\frac{\lambda_k-\lambda_m}{\lambda_k+\lambda_m}.
\ee

Recently Mironov and Morozov \cite{MMQ} conjectured a simple expansion formula for the KW tau-function
\be\label{MMM}
\tau_{KW}= \sum_{\lambda \in \DP} \left(\frac{\hbar}{16}\right)^{|\lambda|/3} \frac{Q_\lambda({\bf t}) Q_\lambda(\delta_{k,1})Q_{2\lambda}(\delta_{k,3}/3)}{Q_{2\lambda}(\delta_{k,1})}.
\ee
Only partitions with the weights divisible by 3 contribute to this expansion.
\begin{remark}
This formula immediately follows \cite{Hp} form Proposition (K') of Di Francesco, Itzykson and Zuber \cite{DFIZ}. Indeed, the main step in its derivation, that is, formula (55) of \cite{MMQ} is nothing but a reformulation of this proposition which uses (\ref{hook}) and the identification of the $f$-functions of \cite{DFIZ} with the Schur Q-functions \cite{JQ}. 
\end{remark}

Let us suggest an analog of the Mironov--Morozov formula for the BGW tau-function:
\begin{conjecture}\label{MC}
\be\label{MF}
\tau_{BGW}= \sum_{\lambda \in \DP} \left(\frac{\hbar}{16}\right)^{|\lambda|} \frac{Q_\lambda({\bf t}) Q_\lambda(\delta_{k,1})^3}{Q_{2\lambda}(\delta_{k,1})^2}.
\ee
\end{conjecture}

\begin{remark}
One can argue that it is much more natural to consider the expansion of these tau-functions in the basis of the Schur functions. For the KW tau-function this expansion was investigated already by Itzykson and Zuber in the early 90's \cite{IZ}. The coefficients of this expansion, which can be described by the determinants of the affine coordinates on the Sato Grassmannian, are rather complicated  \cite{Zhou,BY}, and their relation to geometrical interpretation of the Kontsevich-Witten tau-function is not known yet. This type of expansion for the BGW tau-function is discussed in \cite{ZhouB}.
\end{remark}

Formulas (\ref{MMM}) and (\ref{MF}) are rather surprising. 
Indeed, as discussed in \cite{MMQ}, Schur Q-functions are natural elements of the theory of the BKP hierarchy, but not KdV. In the next section we will show that there is a natural explanation of the appearance of the Schur Q-functions here. Namely, we prove (for generalized BGW model) and conjecture (for KW model), that these tau-functions solve both KdV and BKP hierarchies. 

\begin{remark}
 Another way to relate BKP with KdV (so-called 1-constrained BKP) was described in  \cite{OLax,Cheng}. To the best of our understanding, this reduction does not coincide with the one considered in this paper.
\end{remark}

%%%%%%%%%%%%%%%%%%%%%%%%%%%%%%%%%%%%%%%%%%%

\section{BKP hierarchy}\label{S3}

The BKP hierarchy was introduced by Date, Jimbo, Kashiwara and Miwa in \cite{Date,JM}. 
It can be represented in terms of tau-function $\tau({\bf t})$ by the Hirota bilinear identity, similar to (\ref{HBEKP}),
\be\label{HBE}
\frac{1}{2 \pi i}\oint_{\infty} e^{ \sum_{k\in\z_{\odd}^+}(t_k-t_k') z^k}
\tau ({\bf t}-2[z^{-1}])\tau ({\bf t'}+2[z^{-1}])\frac{dz}{z} =\tau ({\bf t})\tau ({\bf t'}).
\ee
Here
\be
{\bf t}\pm 2[z^{-1}]:= \bigl \{ t_1\pm   
2z^{-1}, 
t_3 \pm \frac{2}{3}z^{-3},t_5 \pm \frac{2}{5}z^{-5}, \ldots \bigr \}.
\ee

\subsection{Symmetries of BKP}\label{BKPS}

Vertex operator 
\be
X(z,w)=\exp\left(\sum_{k\in\z_{\odd}^+}t_k(z^k-w^k)\right)\exp\left(-2\sum_{k\in\z_{\odd}^+}\left(\frac{1}{kz^k}-\frac{1}{kw^k}\right)\frac{\p}{\p t_k}\right)
\ee
generates the (additional) symmetries of the BKP hierarchy \cite{Date}. In particular, 
the {\em Heisenberg-Virasoro subalgebra} of BKP symmetry algebra is generated by the operators 
\be\label{JB}
\widehat{J}_k^B =
\begin{cases}
\displaystyle{2\frac{\p}{\p t_k} \,\,\,\,\,\,\,\,\,\,\,\, \mathrm{for} \quad k>0},\\[2pt]
%\displaystyle{0}\,\,\,\,\,\,\,\,\,\,\,\,\,\,\,\,\,\,\, \mathrm{for} \quad k=0,\\[2pt]
\displaystyle{-kt_{-k} \,\,\,\,\,\mathrm{for} \quad k<0,}
\end{cases}
\ee
for odd $k$ and Virasoro operators
\be
\widehat{L}_k^B=\frac{1}{2}\sum_{i+j=k} :\widehat{J}_i^B\widehat{J}_j^B:
\ee
for even $k$. Here the bosonic normal ordering puts all $\widehat{J}_m^B$ with positive $m$ to the right of all $\widehat{J}_m^B$ with
negative $m$.
Let us also consider the $W^{(3)}$-algebra, which is generated by 
\be
\widehat{M}_k^B=\frac{1}{3}\sum_{i+j+l=k} :\widehat{J}_i^B\widehat{J}_j^B\widehat{J}_l^B:\,\,\,\,\,\,\, k \in {\mathbb Z}_{\odd}.
\ee
Below we will need two of these generators, namely
\begin{equation}
\begin{split}
\widehat{M}_{-1}^B&=\sum_{k,m\in \z_{\odd}^+}^\infty \left(2 kmt_{k}t_{m}\frac{\p}{\p t_{k+m-1}}+4(k+m+1)t_{k+m+1}\frac{\p^2}{\p t_k \p t_m}\right),\\
\widehat{M}_{-3}^B&=\sum_{k,m\in \z_{\odd}^+}^\infty \left(2 kmt_{k}t_{m}\frac{\p}{\p t_{k+m-3}}+4(k+m+3)t_{k+m+3}\frac{\p^2}{\p t_k \p t_m}\right)+\frac{t_1^3}{3}.
\end{split}
\end{equation}

These generators are similar to the cut-and-join operators (\ref{CAJ}) and (\ref{CAJN}).
Namely, after a transformation $t_k \mapsto t_k/2$ we have
\begin{equation}
\begin{split}
\widehat{W}_0(N)&=\sum_{k,m\in \z_{\odd}^+}^\infty \left(\frac{1}{2}kmt_{k}t_{m}\frac{\p}{\p t_{k+m-1}}+(k+m+1)t_{k+m+1}\frac{\p^2}{\p t_k \p t_m}\right)+\left(\frac{1}{16}-\frac{N^2}{4}\right)t_1,\\
\widehat{W}_1&=\frac{1}{3}\sum_{k,m\in \z_{\odd}^+}^\infty \left(\frac{1}{2}kmt_{k}t_{m}\frac{\p}{\p t_{k+m-3}}+(k+m+3)t_{k+m+3}\frac{\p^2}{\p t_k \p t_m}\right)+\frac{t_1^3}{2^3 3!}+\frac{t_3}{16}.
\end{split}
\end{equation}
Therefore we can identify
\begin{equation}
\begin{split}
\widehat{W}_0(N)&=\frac{1}{4}\widehat{M}_{-1}^B+\left(\frac{1}{16}-\frac{N^2}{4}\right)t_1,\\
\widehat{W}_1&=\frac{1}{12}\widehat{M}_{-3}^B-\frac{t_1^3}{2^4\cdot3^2}+\frac{t_3}{16}.
\end{split}
\end{equation}
We see that $\widehat{W}_0(N)$ for any $N$ belongs the algebra of BKP symmetries. Then the cut-and-join representation (\ref{cajg}) implies 
\begin{theorem}\label{MT}
Tau-function of the generalized BGW model $\tau_{BGW}({\bf t}/2,N)$ is a tau-function of the BKP hierarchy.
\end{theorem}

\begin{remark}
Let us note that one can construct another two-parameter family of the solutions of the BKP hierarchy, namely
\be
\tau({\bf t})=\exp\left(\hbar\left(\widehat{W}_1+\frac{t_1^3}{2^4\cdot3^2}+b t_3\right)\right)\cdot 1,
\ee
where $b$ and $\hbar$ are arbitrary parameters.
\end{remark}

\subsection{Hypergeometric solutions of BKP hierarchy}

There are several interesting families of the BKP tau-functions, described by Schur Q-functions.
For example, Schur Q-functions provide all possible polynomial solutions of the BKP hierarchy \cite{Date,You,Nimmo,KL}. 
\begin{remark}
For $N\in {\mathbb Z}+\frac{1}{2}$ the tau-function $\tau_{BGW}({\bf t},N)$ of the generalized BGW model is polynomial, given by shifted a Schur function for a triangular Young tableau \cite{ABGW},
\be
\tau_{BGW}({\bf t},N)\in {\mathbb C}[{\bf t}]\,\,\,\,\,\,\,\,\,\,\,\, \forall N\in {\mathbb Z}+\frac{1}{2}.
\ee
From Theorem \ref{MT} it follows that after rescaling of times, $\tau_{BGW}({\bf t}/2,N)$, these functions  are also solutions of the BKP hierarchy. Hence \cite{KL} they are shifted Schur Q-functions. 
\end{remark}

The class of hypergeometric solutions of the BGW hierarchy was introduced by Orlov \cite{OBKP}. It was shown by Mironov, Morozov, and Natanzon \cite{MMNQ} that they can be interpreted as generating functions of the {\em spin Hurwitz numbers}. Hypergeometric solutions of the BKP hierarchy are given by the following sums over strict partitions
\begin{equation}
\begin{split}\label{HG}
\tau&=\sum_{\lambda \in \DP} 2^{-\ell(\lambda)}r_\lambda Q_\lambda^{{\rm{Mac}}}({\bf t}/2)Q_\lambda^{{\rm{Mac}}}({\bf t}^*/2)\\
&=\sum_{\lambda \in \DP} r_\lambda Q_\lambda({\bf t}/2)Q_\lambda({\bf t}^*/2),
\end{split}
\end{equation}
where
\be
r_\lambda=\prod_{j=1}^{\ell(\lambda)} r(1)r(2)\dots r(\lambda_j)
\ee
for some $r(z)$. It is also convenient to introduce alternative parametrization by $\xi(n)$ for $n \in {\mathbb Z}^+$ with
\be
e^{\xi(n)}:=\prod_{j=1}^n r(j).
\ee
In this parametrization $r_\lambda=e^{\sum_{j=1}^{\ell(\lambda)}\xi(\lambda_j)}$.

We conjecture that the KW and generalized BGW tau-functions belong to this family. These conjectures are implied by the expansions (\ref{MMM}) and (\ref{MF}). 
Let us start from the generalized BGW model. 
\begin{conjecture}\label{C2}
The partition function of the generalized BGW model in the properly normalized times, $\tau_{BGW}({\bf t}/2, N)$ is a hypergeometric tau-function of the BKP hierarchy (\ref{HG}), described by
\be
r(z)=\hbar \frac{(2z-1)^2-4N^2}{16},
\ee
and  $t_k^*=2\delta_{k,1}$. 
\end{conjecture}
Let us stress that for  the BGW model ($N=0$) this conjecture  follows directly from conjecture \ref{MC}. Namely, from (\ref{hook}) we have
\be
\frac{Q_\lambda(\delta_{k,1})}{Q_{2\lambda}(\delta_{k,1})}=\prod_{j=1}^{\ell(\lambda)} (2\lambda_j-1)!!.
\ee

To describe the KW tau-function let us introduce
\be
A_k=\prod_{j=1}^k \frac{(6j-1)(6j-5)}{16}.
\ee
Then for arbitrary $\beta\neq 0$ we define
\begin{equation}\label{RKW}
\begin{split}
e^{\xi_{KW}(3k)}&=\hbar^kA_k,\\
e^{\xi_{KW}(3k-1)}&=-\hbar^{k-1/3}\frac{2}{(6k-1)\beta}A_k,\\
e^{\xi_{KW}(3k-2)}&=\hbar^{k-2/3}\frac{8\beta}{(6k-1)}A_k.\\
\end{split}
\end{equation}
\begin{conjecture}\label{C3}
The KW tau-function in the properly normalized times, $\tau_{KW}({\bf t}/2)$ is a hypergeometric tau-function of the BKP hierarchy, described by
(\ref{HG}) with $ r^{KW}_\lambda=e^{\sum_{j=1}^{\ell(\lambda)}\xi_{KW}(\lambda_j)}$ given by (\ref{RKW}) and $t_k^*=\frac{2}{3}\delta_{k,3}$.
\end{conjecture}
This conjecture and the result of Mironov and Morozov \cite{MMQ} imply a relation
\be
 r^{KW}_\lambda Q_{\lambda}(\delta_{k,3}/3) =\left(\frac{\hbar}{16}\right)^{|\lambda|/3} \frac{ Q_\lambda(\delta_{k,1})}{Q_{2\lambda}(\delta_{k,1})} Q_{2\lambda}(\delta_{k,3}/3)  \,\,\,\,\,\,\,\,\,\,\,\, \forall \lambda \in \DP.
\ee
Let us remind the reader that $Q_{\lambda}(\delta_{k,3}/3)$ vanishes for all partitions with the weight not divisible by 3. Moreover, it also vanishes for some partitions with the weight divisible by 3.

In Conjectures \ref{C2} and \ref{C3} the variables $t_k^*$ are taken at the points, associated with the corresponding dilaton shifts.

\begin{remark}
With John Stembridge's Maple packages SF and QF \cite{SF} we have checked all conjectures perturbatively for $|\lambda|\leq 39$. 
\end{remark}

Conjectures \ref{C2} and \ref{C3} rise numerous questions. In particular, if correct, they lead to the relatively simple interpretation of the intersection theory on the moduli spaces by spin Hurwitz numbers (see \cite{MMNQ} and references therein). Moreover, restoring second set of times $t^*$ one can consider natural deformations to the 2-component BKP hierarchy, which should describe a family of double spin Hurwitz numbers. Geometric interpretation of these families is not known yet. Let us also note that using Conjectures \ref{C2} and \ref{C3} one can find explicit description of the $\tau_{KW}({\bf t})$ and $\tau_{BGW}({\bf t}, N)$ in terms of neutral fermions \cite{OBKP,HLO}.  These topics will be considered elsewhere.

\subsection{Miller--Morita--Mumford classes}

With the forgetful map $\pi: \overline{\mathcal M}_{g,n+1} \rightarrow \overline{\mathcal M}_{g,n} $ we define the {\em Miller--Morita--Mumford tautological classes} \cite{Mumford}, $\kappa_k:= \pi_* \psi_{n+1}^{k+1} \in H^{2k}(\overline{\mathcal{M}}_{g,n},\mathbb{Q})$. According to Manin and Zograf \cite{MZ}, insertion of these classes can be described by the translation of the times $t_k$ responsible for  the insertion of the $\psi$-classes. Translations are symmetries of BKP hierarchy, generated by (\ref{JB}), hence from Theorem \ref{MT} we have
\begin{corollary}
Let us consider the generating function of the higher $\Theta$-Weil--Petersson volumes
\be
\tau_{GWP}^\Theta({\bf t},{\bf s}):=\exp\left(\sum_{g=0}^\infty \sum_{n=0}^\infty \hbar^{2g-2+n}{\mathcal F}_{g,n}({\bf t},{\bf s})\right),
\ee
where
\be
{\mathcal F}_{g,n}({\bf t},{\bf s})=\sum_{a_1,\ldots,a_n\geq 0} \int_{\overline{\mathcal{M}}_{g,n}} \Theta_{g,n}e^{\sum_{k=1}^\infty s_k \kappa_k } \psi_1^{a_1} \psi_2^{a_2}\dots \psi_n^{a_n} \frac{\prod (2 a_i +1)!! t_{2a_i+1}}{n!}.
\ee
For arbitrary values of the parameters ${\bf s}$ the function $\tau_{GWP}^\Theta({\bf t}/2,{\bf s})$ is a tau-function of the BKP hierarchy in variables ${\bf t}$.
\end{corollary}

From Conjecture \ref{C3} a direct analog of this corollary follows for the generating function $\tau_{GWP}({\bf t}/2,{\bf s})$ for the case without Norbury's $\Theta$-classes.

%%%%%%%%%%%%%%%%%%%%%%%%%%%%%%%%%%%%%%%%%%%

\section*{Acknowledgments}
The author is grateful to A. Mironov, A. Morozov, and A. Orlov for useful discussions.  This work was supported by IBS-R003-D1.

\end{document}